\documentclass[a4paper,12pt]{article}
\usepackage[utf8]{in
    putenc}
\usepackage{cancel}
\usepackage{ulem}
\usepackage{amsfonts}
\usepackage{amssymb}
\usepackage{graphicx}
\usepackage{amsmath}
\usepackage{enumerate}
\usepackage{amsmath}
\usepackage{subfloat}
\usepackage{subfig}
\usepackage{color}
\usepackage{float}
\usepackage{here}
\usepackage{cite}
\usepackage{mathrsfs}
\usepackage{float,epsfig}
\usepackage{dcolumn}

\usepackage{graphicx}
\usepackage{bm}
\usepackage{amsmath,amssymb,amsthm}
\usepackage[colorlinks=true,linkcolor=blue]{hyperref}
\hypersetup{citecolor=magenta}
\title{ {\bf    Phase Transition of AdS Black hole in Massive Gravity Revisited via New
Prescription }}
\author{M. Chabab$^{1}$, H. El Moumni$^{2}$, S. Iraoui$^{1}$, K. Masmar$^{3}$\footnote{karima.masmar@edu.uca.ma (Corresponding author)}\\
	\\ 
	{\small $^{1}$ High Energy and Astrophysics Laboratory, Physics Department, FSSM, Cadi Ayyad University,
	}\\
         {\small  Marrakech, Morocco.
	}\\
	{\small $^{2}$ EPTHE, Physics Department, Faculty of Science,  Ibn Zohr University, Agadir, Morocco. }\\
	{\small $^{3}$ Laboratory of  High Energy Physics and Condensed Matter 
HASSAN II University} \\{\small Faculty of Sciences Ain Chock, Casablanca, Morocco. }
}

\date{}
\begin{document}
 \maketitle
 \abstract{
We apply the new thermodynamical method proposed initially by Hindi. et al \cite{base} to revisit the phase structure of the AdS black hole in the massive gravity framework.  In comparison to the standard thermodynamical method used in literature \cite{mass}, we show that this new technical approach can easily reproduce the critical behavior inherent to this black hole configuration and addresses some of the shortcomings of the standard method. We also show the ability  of this new perception  to find   all possible critical points and regions in which phase transitions occur, an impossible task in the standard method.}
 \vspace{-.5cm}
  \tableofcontents
\section{Introduction}

Since the pioneering work by Hawking and Page \cite{hawking}, the connexion of the black hole physics with thermodynamics has gained much attention,  especially from the extended phase space point of view \cite{Dolan:2011xt}. In this framework,  the cosmological constant is regarded as thermodynamic pressure and the black hole mass is identified to the enthalpy. Within these peculiar developments, a beautiful and deep analogy between AdS black holes and van der Waals liquid-gas system has been revealed \cite{KM,our,our1,our2,our3,our4,our5,our6,holo,moiplb,holo1,holo2,our7,our8,Zou:2017juz,Chabab:2017xdw,Chabab:2018zix} opening a new window towards our understanding of strong gravity and black holes physics.

In this context, a great deal of interest has recently been devoted to the modified theories of gravity, such massive gravity \cite{mass,mass1,mass2,mass3,mass4,R1,R2,R3,R4,R5}. The latter theories spectrum show up a massive spin-2 graviton, providing an explanation of the accelerated expansion of the universe without
invoking dark energy. Recently, the observation of the gravitational waves by LIGO collaboration has constrained the graviton mass and set up an upper limit of $m_g\leq1.2\times 10^{-22} eV$ \cite{TheLIGOScientific:2016src}.
Hence, the massive gravities and graviton can  
help to shed light on the foreseen quantum behavior of
gravity \cite{Vasiliev:1995dn,Dvali:2000xg}.

Nowadays many approaches have been raised to probe the black hole phase transitions, ranging from the thermodynamics technics via the localisation of discontinuities of the heat capacity \cite{our1,our2,our3,our4}, thermodynamics geometries by the singularities of their Ricci scalars  \cite{our5,our6}, the AdS/CFT tools including entanglement entropy and two-point correlation function \cite{holo,moiplb,holo1,holo2,Zeng:2015tfj}, to  quasinormal modes\cite{our7,our8,Zou:2017juz} and chaos structure \cite{Chabab:2018lzf}.

 In addition to these standard methods, a new theoretical prescription has recently been proposed \cite{base} based on a new typical equation of state which is originating from the slope of the temperature versus entropy. This new prescription proved to be simple and efficient to disclose the critical behavior of a thermodynamical system while avoiding the deficiencies of the usual methods.	 In general in the standard thermodynamical method we have to deal with high order polynomial functions and looking for their roots, this difficulty can be also found when looking for the singularities of the Ricci scalar in the geothermodynamics formalism, predominantly the denominator of such quantity is a complex formula found by second derivatives calculation. In AdS/CFT technical  background we have to solve a very hard differential equation recalling a numerical treatment, the same constatation can be noted in the quasinormal modes when the critical behavior is associated with the change of slope of frequencies which usually not evident to be observed numerically or graphically.

More explicitly, the investigation of the thermodynamical behavior of a system is based on the study of its equation of state $P=(T,V)$. In the black hole formalism, the equation of states is an equation relates the temperature to pressure (cosmological constant), volume (event horizon radius) and some other black hole parameters $T=T(r_h,\Lambda,Q, \text{ other hairs})$. While the critical point is probed by inflection point definition $\left(\frac{\partial P}{\partial V}\right)_T=\left(\frac{\partial^2}{\partial V^2}\right)_T=0$ determining the possible phase transitions. This method very much depends on the temperature expression and on the number of the parameter on which depend limiting its efficiency for black
holes with non-spherical horizons in most gravitational
theories. To fix such an issue, we should use an alternative
approach to disclose the critical coordinates values in the extended phase space.

Our aim in this work is to revisit  in an easy manner the phase transition and criticality phenomena in the extended phase space of anti-de Sitter black holes in the canonical ensemble of massive gravity using this new approach \cite{base}.

The outline of the rest of the work is as follows: In the next section,
we first present a brief review of the new approach, then we study the essentials of the thermodynamical behavior of the AdS black hole in massive gravity background which presents a van der Waals-like phase picture.  In section $3$, we show that these results can be derived more easily using the new prescription. More precisely, we use the new technic to reveal the coordinates of the critical point for the four-dimensional charged and five-dimensional neutral black hole solutions and prove that they are consistent with those obtained in \cite{mass}. Furthermore, we initiate the analysis of general higher dimension solutions. The last section is devoted to our conclusion.
 
\section{Setup and a brief review of  the method}\label{secp}

A modification of the ideal gas law was proposed by van der Waals to take into account molecular size and molecular interaction forces. It is usually referred to as the van der Waals equation of state \cite{KM,our}:
\begin{equation}
\left( P+\frac{a}{v^{2}}\right) \left( v-b\right) =k_BT(P,v),
\label{Van}
\end{equation}
where, $P$ and $T$ are the pressure and temperature,
respectively, while $v $ is the specific volume.
The constants $a$ and $b$ have positive values and are characteristic of the individual gas. The van der Waals equation of state approaches the ideal gas law $PV=nRT$ as the values of these constants tend to zero. The constant $a$ and $b$ represent corrections for the intermolecular forces and finite molecular size respectively. The critical behavior is governed by the  coordinates of inflexion point.  Mathematically speaking, the latter are  solutions of the following system of equations involving the second derivative calculation of Eq.\eqref{Van}:
\begin{equation}
\left( \frac{\partial P}{\partial v}\right) _{T}=\left( \frac{\partial ^{2}P%
}{\partial v^2}\right) _{T}=0.  \label{inf}
\end{equation}
A straightforward calculations show that the critical coordinates are, 
\begin{equation}
v_{c}=3b,~~~\&~~~P_{c}=\frac{a}{27b^{2}},~~~\&~~~T_{c}=\frac{8a}{27bk_B}.
\label{Critx}
\end{equation}
One can also check that the enthalpy of the van der Waals system are given by \cite{KM}
\begin{eqnarray}
H&=&  \frac{3 k_B T}{2}-\frac{a}{v}+Pv\label{HV}
\end{eqnarray}
where $\phi$ is a constant characterizing the gas and $S$ stands for the entropy,
\begin{equation}
S=k_B\left( \frac{5}{2}+\ln \left[ \frac{v-b}{\phi
}T^{\frac{3}{2}}\right] \right) .
\end{equation}

Since both the entropy and enthalpy depend both on the specific volume $v$ , we can then write,
\begin{equation}\label{xxxx}
\frac{d^{2}H}{dS^{2}}=\left( \frac{dS}{dv}\right) ^{-1}\frac{d}{dv}\left[
\left( \frac{dH}{dv}\right) \left( \frac{dS}{dv}\right) ^{-1}\right].
\end{equation}
and describe the system by the following equation \begin{equation}\label{newx}
\frac{d^{2}H}{dS^{2}}=\frac{b-v}{k^{2}v^{3}}\left[ a\left( 2b-v\right)
+Pv^{3}\right] .
\end{equation}
from which we generate a new equation of state,
\begin{equation}
P_\text{new}=\frac{v-2b}{v^{3}}a.  \label{PNV}
\end{equation}
The critical point coordinates are then easily localized by  considering the extremum of Eq.\eqref{PNV}  involving just the first derivative,
\begin{equation}
v_{c}=3b,~~~~\&~~~~P_{c}=\frac{a}{27b^{2}},
\end{equation}
which are exactly the same as those obtained previously Eq.\eqref{Critx}. As a consequence, we can re-express the  the temperature and enthalpy in the new forms,

\begin{eqnarray}
T_\text{new} &=&\frac{2a\left( b-v\right) ^{2}}{kv^{3}}, \\
H_\text{new} &=&\frac{3 k_B T}{2}-\frac{2 a b}{v^2}
\end{eqnarray}

To summarize, using these new thermodynamical quantities, one can derive in a simple and rigorous way the critical temperature  and enthalpy of any thermodynamical system by recalling  just the extremum calculation. Now, after discussing the main features of this new prescription, we will apply it to the AdS black hole in massive gravity in order to re-investigate its phase transition structure  and show the simplicity of this method.
  \section{  Black hole thermodynamics in massive gravity framework }
  First, we proceed with a brief review of some basic thermodynamic properties of
 AdS black hole in massive gravity. To this end we 
 consider the following action for an $(n+2)$-dimensional massive gravity \cite{mass},
\begin{equation}\label{action}
\mathcal{I}=\frac{1}{16\pi}\int
d^{n+2}x\sqrt{-g}\Big[R+\frac{n(n+1)}{l^2}-\frac{1}{4}F^2+m^2\sum_{i=1}^4c_i\mathcal{U}_i(g,f)\Big],
\end{equation}
where $R$ is the scalar curvature, $l$ is the AdS radius and  $F$ denotes the electromagnetic gauge filed. Here $m$ is the mass term and $f$ is a fixed symmetric tensor
called the reference metric. Furthermore, the suitable constants for massive gravity are dubbed  $c_i$, and the symmetric polynomials of the eigenvalues of the
 ($n+2$)$\times$($n+2$) matrix
$\mathcal{K}^{\mu}_{~\nu}\equiv\sqrt{g^{\mu\alpha}f_{\alpha\nu}}$ denoted by $\mathcal{U}_i$ are:
\begin{eqnarray}
&~&\mathcal{U}_1=[\mathcal{K}],\nonumber\\
&~&\mathcal{U}_2=[\mathcal{K}]^2-[\mathcal{K}^2],\nonumber\\
&~&\mathcal{U}_3=[\mathcal{K}]^3-3[\mathcal{K}][\mathcal{K}^2]+2[\mathcal{K}^3],\nonumber\\
&~&\mathcal{U}_4=[\mathcal{K}]^4-6[\mathcal{K}^2][\mathcal{K}]^2+8[\mathcal{K}^3][\mathcal{K}]+3[\mathcal{K}^2]^2-6[\mathcal{K}^4].
\end{eqnarray}
The square root in $\mathcal{K}$ stands for the matrix square
root, i.e.,
$(\sqrt{A})^{\mu}_{~\nu}(\sqrt{A})^{\nu}_{~\lambda}=A^{\mu}_{~\nu}$,
while the rectangular brackets represent the traces, 
$[\mathcal{K}]=\mathcal{K}^{\mu}_{~\mu}$.

By minimizing the action given by Eq.~\eqref{action}, one can find  a static black hole solution with the spacetime and reference metrics respectively given by,
\begin{equation}
ds^2=-f(r)dt^2+f^{-1}(r)dr^2+r^2h_{ij}dx^idx^j,
\end{equation}
\begin{equation}\label{refmetric}
f_{\mu\nu}=\mathrm{diag}(0,0,c_0^2h_{ij}),
\end{equation}
Here $c_0$ is a positive constant,  $h_{ij}dx^idx^j$ reads as the line
element for an Einstein space with constant curvature $n(n-1)k$, with the values of $k=1, 0, -1 $ correspond to a spherical, flat, or hyperbolic topology of
the black hole horizon, respectively. Based on  the reference
metric of Eq.\eqref{refmetric}, one can write \cite{mass}
\begin{eqnarray}\label{utermeincoor}
&~&\mathcal{U}_1=nc_0/r,\nonumber\\
&~&\mathcal{U}_2=n(n-1)c_0^2/r^2,\nonumber\\
&~&\mathcal{U}_3=n(n-1)(n-2)c_0^3/r^3,\nonumber\\
&~&\mathcal{U}_4=n(n-1)(n-2)(n-3)c_0^4/r^4.
\end{eqnarray}\begin{eqnarray}
f(r)&=&k+\frac{16\pi P}{(n+1)n}r^{2}-\frac{16\pi M}{n V_{n}r^{n-1}}+\frac{(16\pi Q)^{2}}{2n(n-1)V_{n}^{2}r^{2(n-1)}} \nonumber\\\
& &+\frac{c_{0}c_{1}m^{2}}{n}r+c_{0}^{2}c_{2}m^{2}+\frac{(n-1)c_{0}^{3}c_{3}m^{2}}{r}+\frac{(n-1)(n-2)c_{0}^{4}c_{4}m^{2}}{r^{2}}
\end{eqnarray}
where $V_n$ is the volume of space spanned by coordinates $x^i$, $M$
represents the mass of the black hole and  $Q$ is related to its charge. Under the extended phase space proposal  the AdS radius
$l$ is linked to the pressure by the following relation
\begin{equation}P=\frac{n(n+1)}{16\pi l^2}\end{equation}
In this context, the enthalpy $H$, identified to  the black hole mass $M$, can be determined by solving the equation $f(r_h)=0$, so:
\begin{eqnarray}\label{enthalpy}
H=M&=&\frac{nV_nr_h^{n-1}}{16\pi}\Big[k+\frac{16\pi
P}{(n+1)n}r_h^2+\frac{(16\pi Q)^2}{2n(n-1)V_n^2r_h^{2(n-1)}}+\frac{c_0c_1m^2}{n}r_h+c_0^2c_2m^2\nonumber\\
&~&+\frac{(n-1)c_0^3c_3m^2}{r_h}+\frac{(n-1)(n-2)c_0^4c_4m^2}{r_h^2}\Big].
\end{eqnarray}

 where $r_h$ is the event horizon radius. The Hawking
temperature reads as \cite{mass}, 

\begin{eqnarray}\label{Htemp}
T=\frac{1}{4\pi}f'(r_h)&=&\frac{1}{4\pi r_h}\Big[(n-1)k+\frac{16\pi
P}{n}r_h^2-\frac{(16\pi Q)^2}{2nV_n^2r_h^{2(n-1)}}+c_0c_1m^2r_h+(n-1)c_0^2c_2m^2\nonumber\\
&~&+\frac{(n-1)(n-2)c_0^3c_3m^2}{r_h}+\frac{(n-1)(n-2)(n-3)c_0^4c_4m^2}{r_h^2}\Big].
\end{eqnarray}

while the entropy $S$, the thermodynamic volume $V$ which  is the Legendre transform of the pressure,  and the electric potential $\Phi_Q$ are given by,
\begin{equation}
S=\int_{0}^{r_h} T^{-1} \left(\frac{d H}{d r} \right)_{Q,P}  dr=\frac{V_{n}}{4}r_h^{n}
\end{equation}
\begin{equation}
V=\left( \frac{d H}{d r} \right)_{S,Q}=\frac{V_{n}}{n+1}r_h^{n+1}
\end{equation}
\begin{equation}
\Phi=\left(\frac{\partial H}{\partial
Q}\right)_{S,P}=\frac{16\pi}{(n-1)V_nr_h^{n-1}}Q.
\end{equation}

All the previous thermodynamical quantities generally obey to the first law of black hole thermodynamics in the extended phase space,
\begin{eqnarray}\label{firstlaw}\nonumber
\mathrm{d}H&=&T\mathrm{d}S+V\mathrm{d}P+\Phi\mathrm{d}Q+\frac{V_nc_0m^2r_h^n}{16\pi}\mathrm{d}c_1+\frac{nV_nc_0^2m^2r_h^{n-1}}{16\pi}\mathrm{d}c_2+\frac{n(n-1)V_nc_0^3m^2r_h^{n-2}}{16\pi}\mathrm{d}c_3\\&+&\frac{n(n-1)(n-2)V_nc_0^4m^2r_h^{n-3}}{16\pi}\mathrm{d}c_4,
\end{eqnarray}
where the coupling constants $c_i$ are introduced as thermodynamical variables. The corresponding Smarr formula can be retrieved via scaling argument as,
\begin{eqnarray}\nonumber
(n-1)H&=&nTS-2PV+(n-1)\Phi
Q-\frac{V_nc_0c_1m^2}{16\pi}r_h^n+\frac{n(n-1)V_nc_0^3c_3m^2}{16\pi}r_h^{n-2}\\&+&\frac{n(n-1)(n-2)V_nc_0^4c_4m^2}{8\pi}r_h^{n-3}.
\end{eqnarray}

Having shown the relevant thermodynamical quantities associated   to the  phase transition picture, we would like to revisit,  in the subsequent subsections, the phase structure via the  new prescription presented in Sec.~\ref{secp}, especially for four,  five dimensional black hole solutions and we will comment briefly the high dimensions case.

\subsection{Four dimensional charged black holes}
In this subsection, we will deal with a black hole solution in the $4$-dimensional spacetime background, where for simplicity we consider the constants $c_3$ and $c_4$ equal to $0$. In this case, the enthalpy is reduced in four dimensional spacetime to
 \begin{equation}\label{H}
H=\frac{V_2r_h}{8\pi}\Big[k+\frac{8\pi P}{3}r_h^2+\frac{(8\pi
Q)^2}{V_2^2r_h^2}+\frac{c_0c_1m^2}{2}r_h+c_0^2c_2m^2\Big].
\end{equation}
Under these assumptions and thanks to the Hawking temperature in Eq.\eqref{Htemp},  the black hole
 equation of state reads as, 
\begin{equation}\label{P}
P=\Big(\frac{T}{2}-\frac{c_0c_1m^2}{8\pi}\Big)\frac{1}{r_h}-\Big(\frac{k}{8\pi}+\frac{c_0^2c_2m^2}{8\pi}\Big)\frac{1}{r_h^2}+\frac{8\pi
Q^2}{V_2^2}\frac{1}{r_h^4},
\end{equation}

from which we determine the critical coordinates of phase transition by   the standard method in which we need to  solve the following system of equations:
\begin{equation}\label{cricondition}
\frac{\partial P}{\partial
r_h}\Big|_{r_h=r_{hc},T=T_c}=\frac{\partial^2P}{\partial
r_h^{~2}}\Big|_{r_h=r_{hc},T=T_c}=0.
\end{equation}

To proceed further with calculations, we first introduce a novel notation for the coefficients shown in
Eq.(\ref{P}),
\begin{eqnarray}\label{w124}
&~&w_1=\frac{T}{2}-\frac{c_0c_1m^2}{8\pi}\nonumber\\
&~&w_2=-\Big(\frac{k}{8\pi}+\frac{c_0^2c_2m^2}{8\pi}\Big)\nonumber\\
&~&w_4=\frac{8\pi Q^2}{V_2^2}.
\end{eqnarray}

Thus, the equations Eq.\eqref{H} and Eq.\eqref{P} reduce to:
 \begin{equation}
H=\frac{r_h V_{2} \left(\frac{1}{2}
   c_{0} c_{1} m^2
   r_h+\frac{8}{3} \pi  P r_h^2+\frac{8
   \pi  \omega_{4}}{r_h^2}\right)}{8 \pi }-r_h
   V_{2} \omega_{ 2}
   \end{equation}
\begin{equation}\label{statep}
P=\frac{\omega_{1}}{r_h}+\frac{\omega_{2}}{r_h^2}+\frac{\omega_{4}}{r_h^4}.
\end{equation}
and the critical point resulting from Eq.(\ref{cricondition}) 
are then given by, 
 \cite{mass}
\begin{equation}\label{Crit}
 r_{hc}=\sqrt{-\frac{6w_4}{w_2}},\ 
 w_{1c}=-\frac{4}{3}w_2\sqrt{-\frac{w_2}{6w_4}}
\ and \ 
 P_c=\frac{w_2^2}{12w_4}
\end{equation}

In order to accommodate  the occurence of the critical behavior, we simply impose the condition $w_2<0$. In addition, on can easily notice a universal relation among critical pressure $P_c$, shifted temperature $w_{1c}$, and the horizon radius $r_{hc}$,
\begin{equation}
\frac{P_cr_{hc}}{w_{1c}}=\frac{3}{8}
\end{equation}

After revisiting the essential of the thermodynamical  critical behavior by recalling the inflexion point proprieties, we call upon the new method in the case of $4d$ charged black hole  to have a clear comparison of the method's efficiency. Since the entropy $S(r_h)$ and  the enthalpy $H(r_h)$ are radius
dependent, we can re-express Eq.\eqref{xxxx} the following form,
\begin{equation}
\frac{d^{2}H}{dS^{2}}=\left(\frac{dS}{dr}\right)^{-1}\frac{d}{dr}\left[\left(\frac{dH}{dr}\right)\left(\frac{dS}{dr}\right)^{-1}\right],\quad with \;\;S=\frac{r_h^2 V_{2}}{4}
\end{equation}


which yields a novel equation of state:
\begin{equation}
\frac{d^{2}H}{dS^{2}}=\frac{4 \left(P r_h^4+r_h^2 \omega
   _{2}+3 \omega _{4}\right)}{r_h^5
 V_{2}}.
\end{equation}
 By solving this equation with respect to $P$, we get a new expression of the pressure  which is totally distinct from the ordinary equation  of state obtained in Eq.\eqref{statep}
\begin{equation}\label{Pnew}
P_{new}=\frac{-r_h^2 \omega _{2}-3.
   \omega _{4}}{r_h^4}
\end{equation}
 At this level, we can obviously see the difference in the complexity of formula related to each equation of state Eq.\eqref{statep} and Eq.\eqref{Pnew} in the high order terms on $r_h$ and one can note that the new equation of state Eq.\eqref{Pnew} is more appropriate to handle.
By using this new formula of the pressure, 
we can show that the temperature, enthalpy can be rewritten as,
\begin{equation}
T_{new}=\frac{c_{0} c_{1} m^2}{4 \pi
   }-\frac{4(r^2\omega_2+2 \omega _{4})}{r_h^3}
\end{equation}

Note that these quantities do not depend on the pressure anymore  implying a reduction in the number of parameters.
\begin{equation}
\omega_{1new}=-\frac{2(r^2 \omega_2+2 \omega _{4})}{r_{h}^3}
\end{equation}
\begin{equation}
H_{new}=\frac{r_h V_{2} \left(3 c_{0} c_{1} m^2 r_h-64 \pi 
 \omega _{2}\right)}{48 \pi },
   \end{equation}

\begin{figure}[h!t]
\includegraphics[scale=.5]{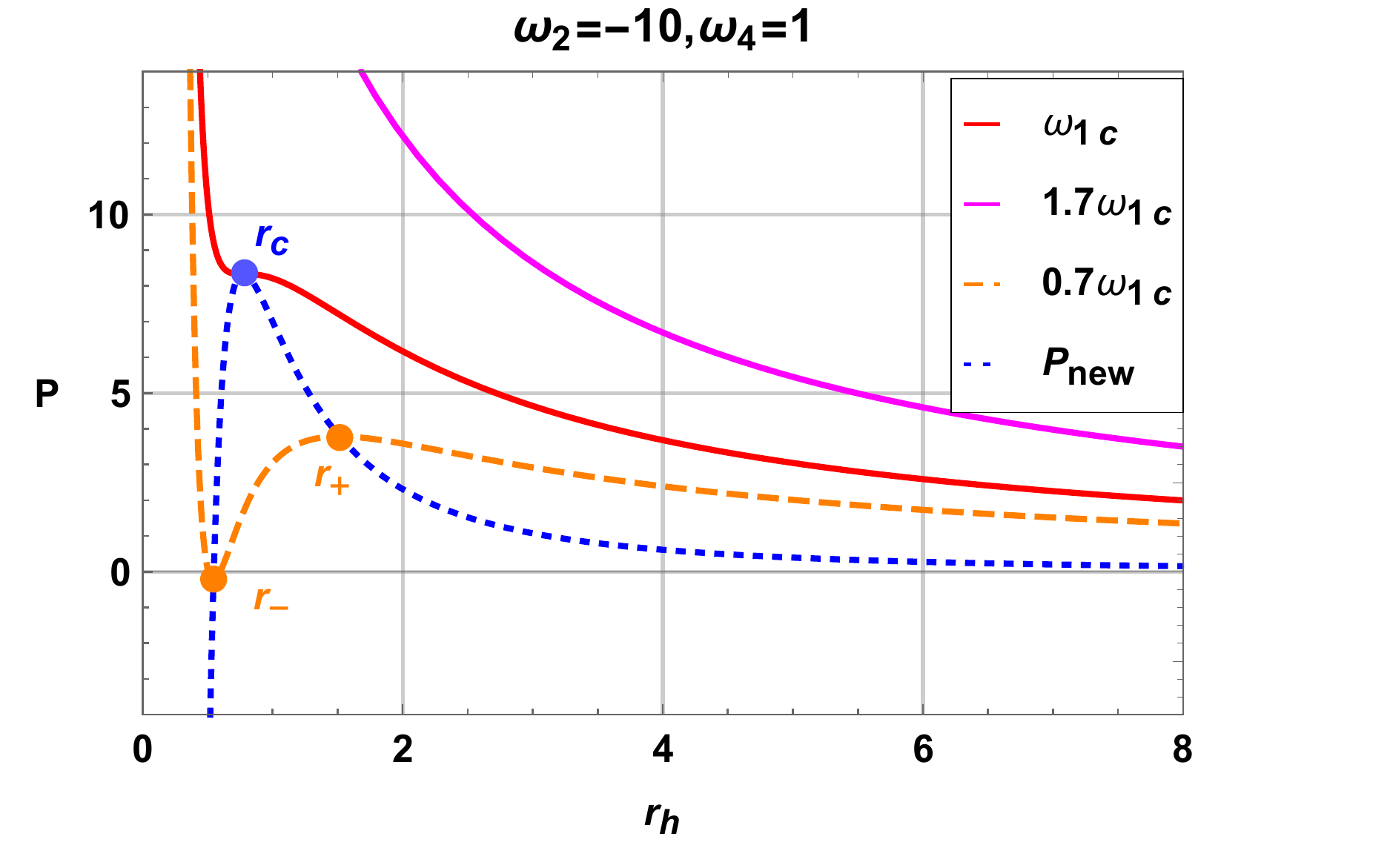}
\includegraphics[scale=.5]{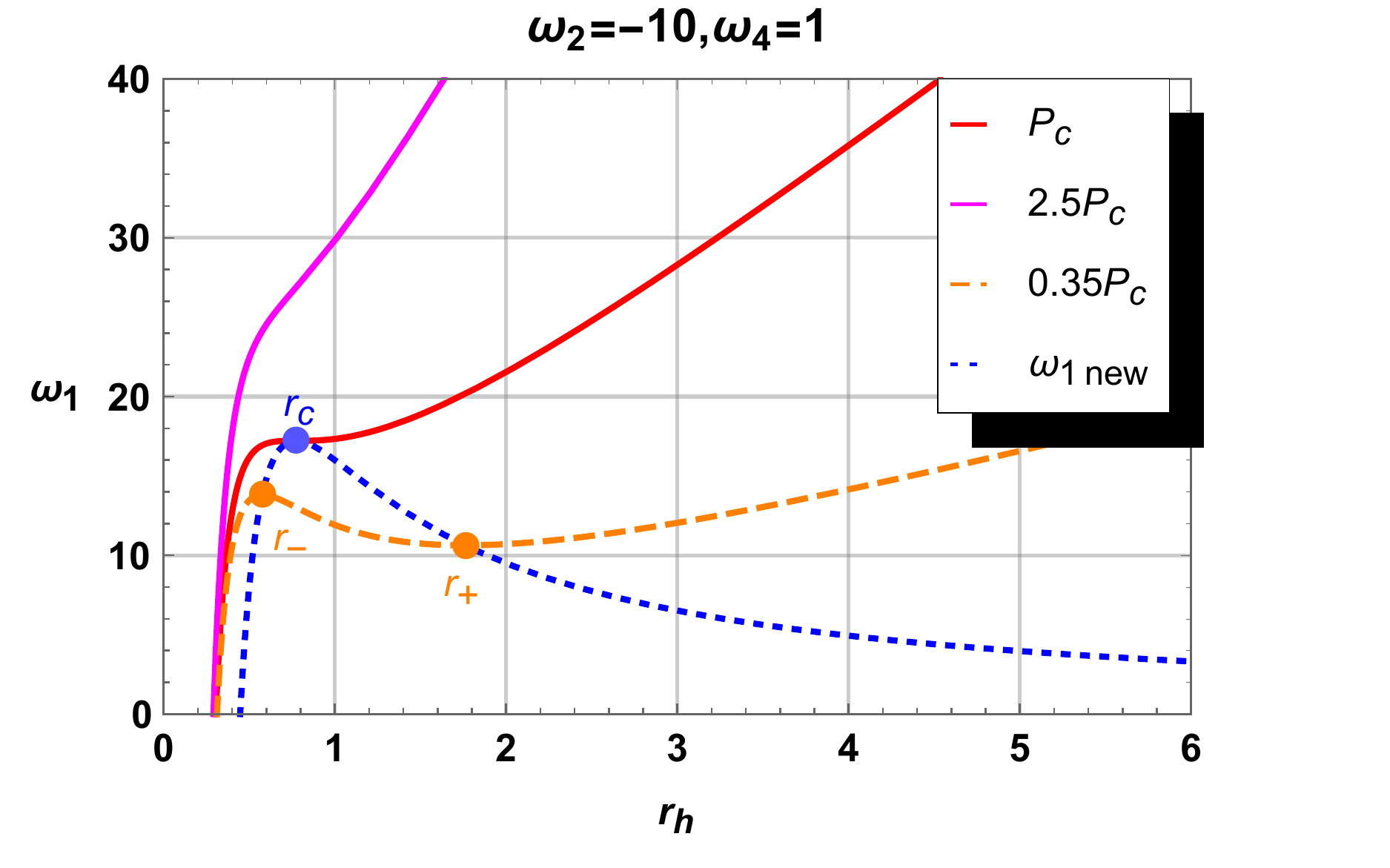}
\caption{The four-dimensional $(P-r_h)$ and  $(\omega_{1}-r_h)$ diagrams  for $w_2=-10$ and $w_4=1$ .}
\label{fig1}
\end{figure}

Then we easily derive the extremum of the equation of state, using  just the vanishing first derivative formula
     \begin{equation}
\left(\frac{dP_{new}}{dr}\right)|_{r=r_{NC}}=0
   \end{equation}
  Its solutions lead to the new critical (NC) coordinates, namely the horizon radius $r_{NC}$, temperature $T_{NC}$ and pressure $P_{NC}$
        \begin{equation}
r_{NC}=\sqrt{\frac{-6\omega_4}{\omega_2}},\quad T_{NC} = \frac{1}{36} \left(\frac{9 c_{0}
  c_{1} m^2}{\pi }+8 \sqrt{\frac{6 \omega_2^3}{-\omega_4}}\right),\quad 
P_{NC} = \frac{\omega _{2}^2}{12
  \omega _{4}}.
   \end{equation}
These quantities are identical to those obtained in Eq.\eqref{Crit} proving the validity of this new approach and its consistency with the usual one of the extended phase space. More than that, one  extract easily all the possible critical points that a system can acquire. In light of this  new method, we plot in Fig.\ref{fig1} the variation of the old and new pressures versus the horizon radius $r_h$ in the left panel, the right one is dedicated  to the shifted temperatures  variation in term of $r_h$ in order to show the perfect agreement with the standard thermodynamical  technic    

From Fig.\ref{fig1}, we can see that 
the $P-r_h$  diagram   depicted  in the left panel is  evidently  the same as the plot of the
van der Waals liquid-gas system.  Therefore, for
$\omega_1<\omega_{1c}$, there is a small-large black hole phase transition corresponding to the liquid-gas phase transition of the van der Waals. Moreover we see that such first order phase transition becomes of second order when $\omega_1=\omega_{1c}$ as in the case of the real-gas system.

For the right panel  temperatures variation  shows that the Hawking
temperature is a monotonic function if  $\omega_1>\omega_{1c}$, but when $\omega_1\leq\omega_{1c}$,  it represents a critical point where the phase transition is a second order one,  which is in many ways reminiscent of the liquid/gas transition of a van der Waals fluid. The  dotted blue line is noting than the so-called saturation line and $r_\pm$ present the intersection of this curve with isobaric/isotherm curve for the both panels.  Between these two value $r_-<rh<r_+$, the black hole system is unstable, there is a phase transition
between a small and a large black hole.  It is worth noting that for, the intersection points meet and are equal to the critical horizon radius. This line represents all the possible critical points, therefore all possible phase transitions, contrary to the standard method in which obtaining all points between which phase transitions occur, one must consider all pressures equal to or smaller than the critical pressure, technically speaking it's an impossible task.

After showing that the new prescription completely agrees with the results derived by the standard method for the four-dimensional charged black hole solution,
and its ability  to map all possible critical points and regions in
which phase transitions take place.
 we will turn our attention to see how this new powerful and easy to use recipe operates in the five dimension neutral black holes, and if it can reveal the critical point coordinates  with the same fluency as in 4d case.  

  \subsection{Five dimensional neutral black holes}
  
  Here, we focus on  the five-dimensional  neutral black hole, by setting $n=3$ and $\mathcal{U}_4=0$. So we can also set $c_4=0$ in the metric function,  because the 
 $c_3m^2$ term in five-dimensional neutral black holes can play a similar role as the charge.  Then, for simplicity,  we deal with vanishing charge case as in \cite{mass}. The enthalpy Eq.~\eqref{enthalpy} becomes under these assumptions, 
\begin{eqnarray}
H&=&\frac{3V_3r_h^2}{16\pi}\Big[k+\frac{4\pi
P}{3}r_h^2+\frac{c_0c_1m^2}{3}r_h+c_0^2c_2m^2+\frac{2c_0^3c_3m^2}{r_h}\Big].\\
&=&\frac{3 r^2 V_{3}
   \left(\frac{1}{3} c_{0}
  c_{1} m^2 r+\frac{4}{3}
   \pi  P r^2\right)}{16 \pi
   }-\frac{3}{2} r^2 V_{3}
   \omega _{2}-3 r
   V_{3} \omega _{3}
\end{eqnarray}
with,
  \begin{equation}
\omega_{3}=-\frac{c_{0}^{3}c_{3}m^{2}}{8\pi}
\end{equation}

and the  equation of state stands for, 
\begin{eqnarray}\label{statep5}
P&=&\Big(\frac{3}{4}T-\frac{3c_0c_1m^2}{16\pi}\Big)\frac{1}{r_h}-\Big(\frac{3k}{8\pi}+\frac{3c_0^2c_2m^2}{8\pi}\Big)\frac{1}{r_h^2}-\frac{3c_0^3c_3m^2}{8\pi}\frac{1}{r_h^3}.\\
& = &\frac{3\omega_{1}}{2r_h}+\frac{3\omega_{2}}{r_h^2}+\frac{3\omega_{3}}{r_h^3}
\end{eqnarray}

From Eq.\eqref{cricondition}, we find the coordinates of the critical point 
 $r_h$, $w_1$, and $P$  as follows:
\begin{equation}
r_{hc}=-\frac{3w_3}{w_2},\ 
w_{1c}=\frac{2w_2^2}{3w_3}, \ 
P_c=-\frac{w_2^3}{9w_3^2}
\end{equation}
By the same token of the  previous section, keeping $w_2<0$ and $w_3>0$ ensure  the existence of  a critical behavior, while the universal number becomes in this case ,
\begin{equation}
\frac{P_cr_{hc}}{w_{1c}}=\frac{1}{2}
\end{equation}
The essential of the thermodynamical quantities has been found using the standard thermodynamical method, we recall  the new technic to recheck its validity. In this context, the  enthalpy still given by Eq.\eqref{xxxx} while the entropy in five dimension reads as,

    \begin{equation}\label{xx1}
  S=\frac{r_h^3 V_{3}}{4}
    \end{equation}
Substituting Eq.\eqref{xx1} in Eq.\eqref{xxxx}  leads to following equation:
\begin{equation}\label{zzzw}
\frac{d^{2}H}{dS^{2}}=\frac{16 \left(P r_h^3+3 r_h
  \omega _{2}+6
  \omega _{3}\right)}{9
   r_h^5 V_{3}}
   \end{equation}
The resolution of Eq.\eqref{zzzw} yields a new formula of the pressure and consequently  the following equation of state,
\begin{equation}\label{Pnew5}
P_{new}=-\frac{3 (r_h \omega
   _{2}+2 \omega
   _{3})}{r_h^3}.
\end{equation}

 Then, by looking for the extremum  $\left(\frac{dP_{new}}{dr}\right)|_{r=r_{NC}}=0$ to determine the critical coordinates, we obtain the new  thermodynamical quantities, 
\begin{equation}
T_{new}=\frac{c_{0} c_{1}
   m^2}{4 \pi }-\frac{4 (2 r_h
  \omega _{2}+3
   \omega _{3})}{r_h^2},
\end{equation}
\begin{equation}
H_{new}=\frac{r_h V_{3}
   \left(c_{0} c_{1}
   m^2 r_h^2-36 \pi  (r_h
  \omega_{ 2}+2
   \omega
   _{3})\right)}{16 \pi },
   \end{equation}
   \begin{equation}
r_{NC}=-\frac{3 \omega
   _{3}}{\omega _{2}}
, \ 
 T_{NC} = \frac{c_{0} c_{1}
   m^2}{4 \pi }+\frac{4
  \omega _{2}^2}{3
   \omega _{3}}, \ 
P_{NC} = -\frac{\omega _{2}^3}{9
  \omega _{3}^2}
     \end{equation}  

\begin{figure}[!ht]
\includegraphics[scale=.5]{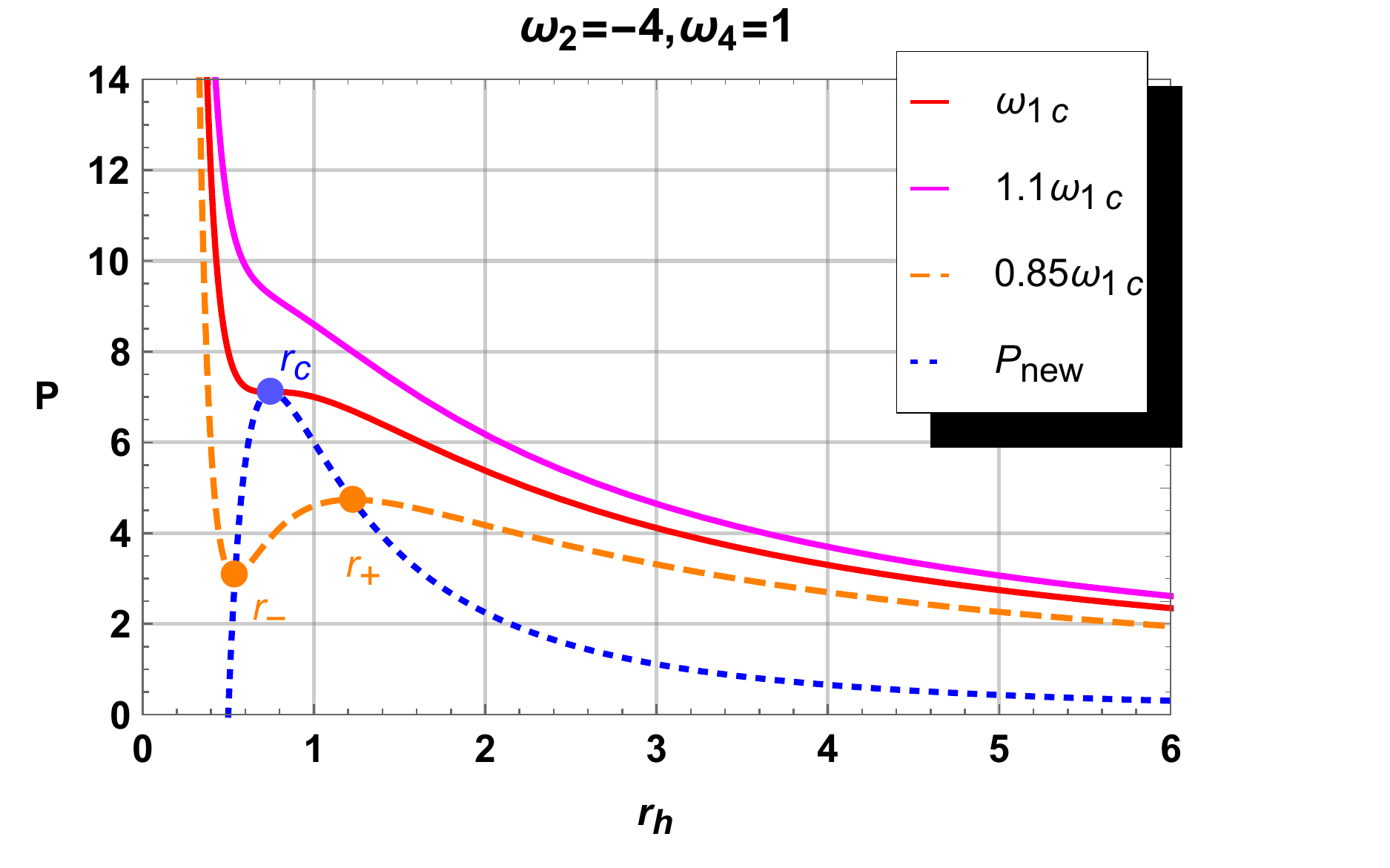}
\includegraphics[scale=.5]{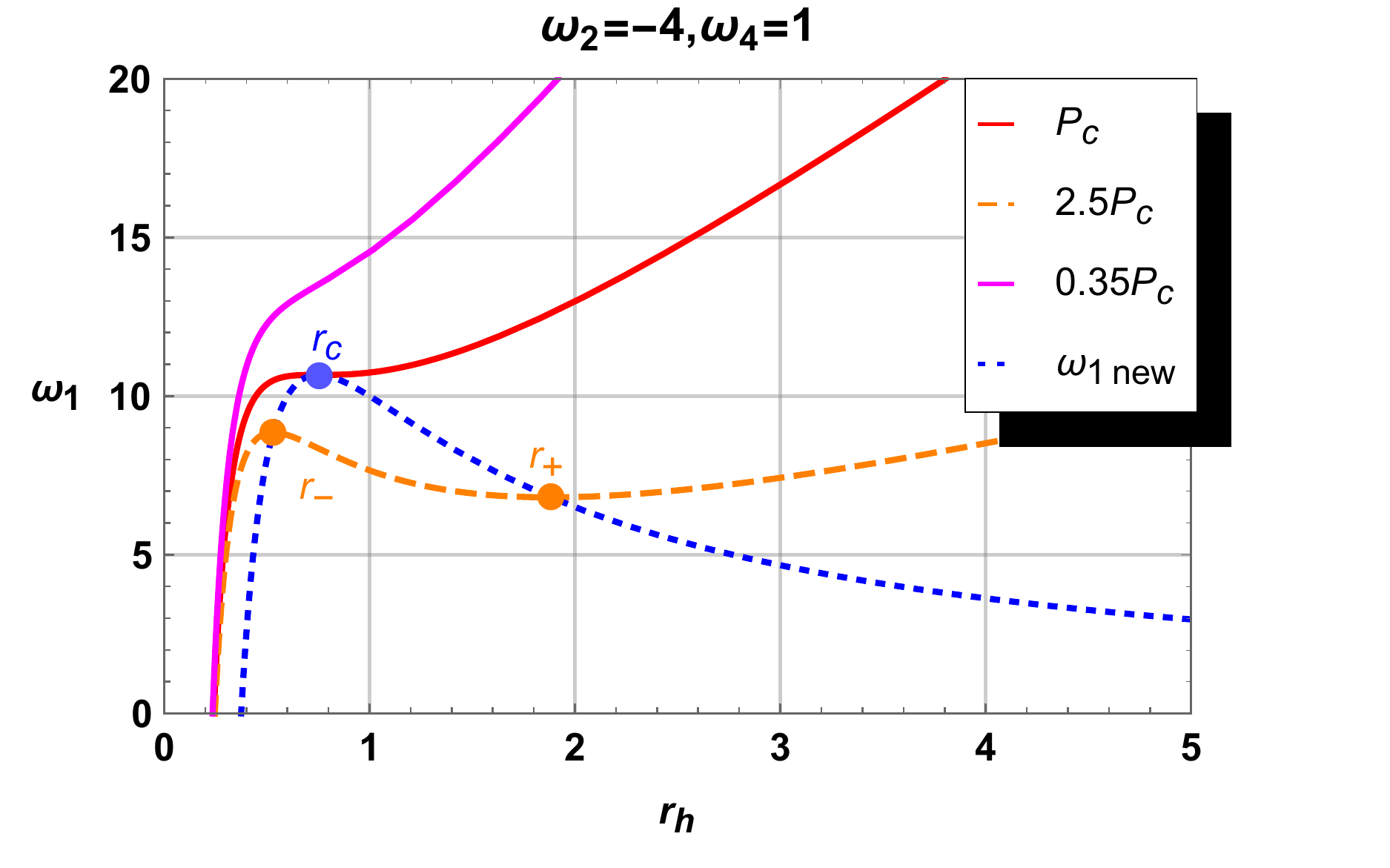}
\caption{The 5-dimensional $(P-r_h)$ and  $(\omega_{1}-r_h)$ diagrams  for $w_2=-4$ and $w_3=1$ .}
\label{fig2}
\end{figure}

   Once again  one can observe the simplicity of the new equation  of states Eq.\eqref{Pnew5} compared to the standard one Eq.\eqref{statep5}, also
   we see that the new method reproduces exactly  the critical quantities. In order to illustrate the validity and the power of this prescription,  we plot  as in the previous section in  Fig.\ref{fig2} the variation of the pressures and  temperatures as function of the horizon radius. 
From which the same description and analysis hold confirming the pertinence  of this approach to probe the phase structure in five dimensional neutral AdS black hole framework. In the end of this section, it is worth to point out some further advantages of this new prescription   
which  provides the possibility to get different thermodynamical quantities independent of each other and to only include the critical points. More precisely, one can find  analytically all possible phase transition horizon radii and their corresponding pressures  by calculating just the first derivative of the state equation.

  \subsection{General higher dimensional black holes}

 To complete our investigation, we will initiate here a study of the black hole solution in an arbitrary dimension and topology.   We recall  the standard equation of state for such charged
black holes which  is given by inverting Eq.\eqref{Htemp} as
\begin{eqnarray}\label{nP}
P&=&\Big(\frac{nT}{4}-\frac{nc_0c_1m^2}{16\pi}\Big)\frac{1}{r_h}-\frac{n(n-1)(k+c_0^2c_2m^2)}{16\pi}\frac{1}{r_h^2}-\frac{n(n-1)(n-2)c_0^3c_3m^2}{16\pi}\frac{1}{r_h^3}\nonumber\\
&~&-\frac{n(n-1)(n-2)(n-3)c_0^4c_4m^2}{16\pi}\frac{1}{r_h^4}+\frac{8\pi
Q^2}{V_n^2}\frac{1}{r_h^{2n}}.
\end{eqnarray}

  In this general background, by using  the full expression of enthalpy given by Eq.\eqref{enthalpy},  Eq.\eqref{xxxx} reads as,
\begin{eqnarray}\nonumber
\frac{d^{2}H}{dS^{2}}&=&\frac{r_h^{-3 (n+1)} }{\pi  n^2 V_n^3}\left((1-n) n V_n^2 r_h^{2 n} \left(3 c_0^4 c_4 m^2
   (n-3) (n-2)+2 c_0^3 c_3 m^2 (n-2) r_h+r_h^2 \left(c_0^2 c_2
   m^2+k\right)\right)\right.\\ &+&\left.16 \pi  P V_n^2 r_h^{2 n+4}+128 \pi ^2 (2 n-1) Q^2
   r_h^4\right)
\end{eqnarray}
 which leads to a totally new equation of  state,
 \begin{eqnarray}\label{pnnew}\nonumber
 P_{new}&=& \frac{3 c_0^4 c_4 m^2 (n-3) (n-2) (n-1) n}{16 \pi  r_h^4}+\frac{c_0^3 c_3 m^2 (n-2) (n-1) n}{8 \pi  r_h^3}\\
 &+& \frac{(n-1) n \left(c_0^2 c_2 m^2+k\right)}{16 \pi  r_h^2}
 +\frac{8 \pi  (1-2 n) Q^2 r_h^{-2 n}}{V_n^2}
 \end{eqnarray}
  
\begin{figure}[!ht]
      \centering
\includegraphics[scale=.5]{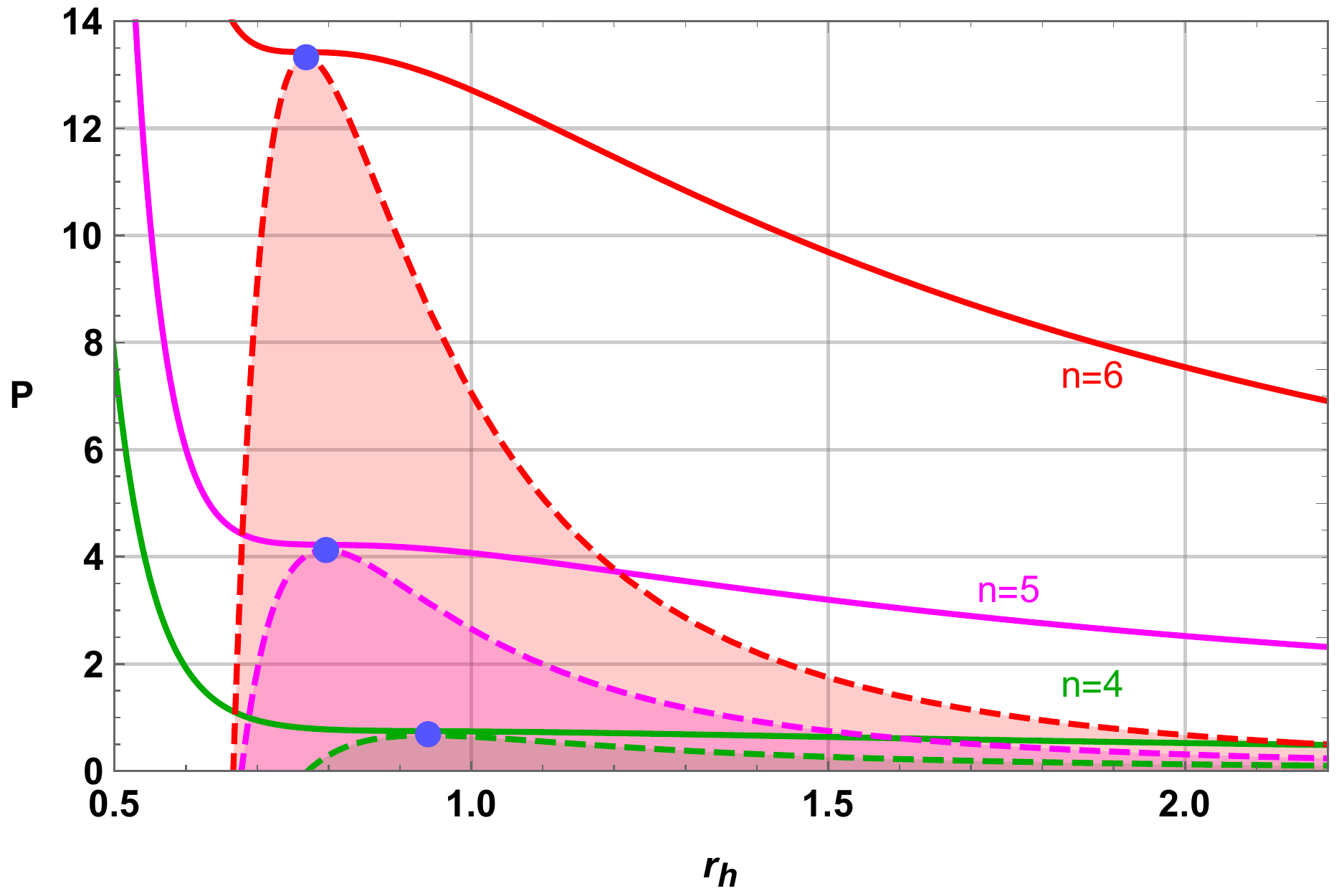}
\caption{The (n+2)-dimensional $(P-r_h)$  diagrams  for $c_{i=\{0\ldots 4\}}=1$ and  $(Q,m)=(1,\frac{1}{2})$.}
\label{fig3}
\end{figure}

From Eq.\eqref{pnnew} we can evidently  see that, apart the charge term, all the $r_h$ powers are independent of spacetime dimension and give the same qualitative contribution to the phase transition in any dimension.  Comparing the new equation of state  and the standard one a blatant remark can be observed is that the absence of the temperature $T$ in the expression $P_{new}$ reducing the number of freedom degrees and simplifying the calculation . Furthermore,  we note that the new expression of the pressure has a polynomial form, where the positive coefficients reflect the repulsion interactions while the negative ones are related to the attractive forces. This new form of the pressure is useful for determining the critical radius via derivation of the roots of Eq.\eqref{pnnew}. However, in view of  the complexity of this equation due to the fact that the pressure is a polynomial of high degrees, its analytical resolution is rather complicated, then a  numerical treatment is more appropriate. As illustration, we depict in Fig.\ref{fig3}, the variation of the pressure in term of the horizon radius using the complete form of the equation of state Eq.\eqref{Htemp} and Eq.\eqref{pnnew},
 
 The dashed lines shown in Fig.\ref{fig3} represent the new plots of the pressure for the values of  $n=4, 5, 6$, hence for $d=6,7,8$. As to the blue points, they display the maximums corresponding to the critical points. For our configuration, the following table summarizes the critical point coordinates corresponding to $6$, $7$ and $8$ dimensions,

\begin{table}[h!t]
\begin{center}\begin{tabular}{|l|l|l|}
 \hline
        dimension  (n+2)     &  $r_{h,c}$ &   $P_c$        \\ \hline\hline
     ${\tt n=4}$  & 0.937647 & 0.667207 \\ \hline\hline
     ${\tt n=5}$  & 0.796435 & 4.12433 \\ \hline\hline
     ${\tt n=6}$  & 0.767787 & 13.3374 \\ \hline\hline
       

  \end{tabular}
\end{center}
\caption{\footnotesize The coordinate of the critical points for some high dimensions.}\label{tab2}
\end{table}

\section{Conclusion}  

In this paper, we have revisited the phase transition of the AdS black hole in the context of the massive gravity by using the new recipe proposed by Hendi. et al. \cite{base}, to show that this black hole system exhibits a van der Waals-like phase structure. This new prescription is based on the slope of the temperature versus the entropy while in the usual the temperature is treated as the equation of state.

More precisely, we have applied this new powerful tool to easily track down the critical coordinates in the four charged and five neutral black hole configurations that were initially derived in \cite{mass}, hence showing a full agreement with the standard thermodynamical methods. In addition, we have initiated a discussion for the case of arbitrary higher dimension, in particular, we presented a brief analysis in the case of $d=6,7,8$ dimensions. \\

In the end, we have proved that this new approach is simple and quite competitive compared to the other technics. Its power resides in the fact that it allows us to easily find the extremum of the first derivative instead of identifying the roots of the second derivative of the equation of state, obtaining new relations for different thermodynamical quantities by eliminating the dependency between such quantities and mapping all possible critical points and regions in
which phase transitions take place, impossible task in the standard method.
Moreover, it can be applied well beyond black hole thermodynamics to other physics areas, such as condensed matter physics and  AdS/CFT correspondence. This approach can be also translate the dynamic of phase transitions \cite{hayat,hayat1} to black hole level.

\end{document}